\documentclass[amssymb,twocolumn,aps,floatfix,showpacs]{revtex4-1}
\usepackage{graphicx}

\usepackage{epstopdf}
\begin{document}
\title{Double Ionisation in R-Matrix Theory Using a 2-electron Outer Region}
\author{Jack Wragg, J. S. Parker, H. W. van der  Hart}

\affiliation{Centre for Theoretical Atomic, Molecular and Optical Physics, School of Mathematics and Physics,
Queen's University Belfast, Belfast, BT7 1NN, United Kingdom}

\begin{abstract}
	We have developed a two-electron outer region for use within R-matrix theory to describe double ionisation processes. The capability of this method is demonstrated for single-photon double ionisation of He in the photon energy region between 80 eV to 180 eV. The cross sections are in agreement with established data. The extended RMT method also provides information on higher-order processes, as demonstrated by the identification of signatures for sequential double ionisation processes involving an intermediate He$^{+}$ state with $n=2$. 
\end{abstract}

\maketitle

The development of laser sources capable of generating ultra-short light pulses or intense XUV and X-ray laser light \cite{Corkum2007,Krausz2009,Yabashi2013} has put additional emphasis on the investigation of multiple ionisation processes \cite{Bergues2012}. To complement recent experimental developments, there is a need for new computational techniques able to describe the multiple ionisation of general atoms. Although progress has been made in the description of double photoionisation of general atoms \cite{Yip2013}, further progress is needed to enable investigation of these processes in a time-dependent manner. This is of particular importance in experiments involving ultra-short light pulses. 

One possible route to develop capability for the description of double-ionisation processes is R-Matrix theory. Time-independent R-matrix theory has already been extended for this purpose. The R-matrix with pseudo-states technique (RMPS) utilises a large basis set in the inner region, which includes residual-ion states in the continuum. Although the first electron can escape, the second must remain bound. Nevertheless, information on double ionisation can be obtained through evaluating the probability that the residual ion is left in an excited state above the ionisation threshold \cite{Bartschat1996}. A second R-matrix approach for double ionisation is intermediate-energy R-matrix (IERM) theory, in which two electrons are allowed to escape the inner region. This leads to a 2-dimensional propagation of the R-matrix over a large distance where the R-matrix is matched to asymptotic solutions. This approach was initially applied to scattering \cite{Burke1987, Scott2009}, and has recently been applied to double photoionisation \cite{McIntyre2013}. 

One promising computational method for solving the time-dependent Schr\"odinger equation (TDSE) for many electron atoms in laser fields is R-Matrix with time-dependence (RMT) \cite{Moore2011, Lysaght2012, Nikolopoulos2008}. With electron-electron interactions fully incorporated, RMT has been used to successfully model a number of features of many-electron atoms undergoing photo-ionisation \cite{Rey2014, Morgan2014}. In RMT, an R-Matrix basis set inner region \cite{burke2011} is attached to an outer-region finite difference grid \cite{Smyth1998} with the ability to describe an ionised electron, enabling the physical processes in each region to be modelled with an appropriate numerical method.

While previous applications of RMT to double-ionisation \cite{Hugo2014} have calculated accurate two-photon cross sections, the range of observables obtainable with this method is narrowed by an upper limit on the radial distance travelled by the inner ionising electron, causing reflections upon interaction with the inner-region boundary. In this communication, we report a method capable of modelling both ejected electrons accurately over a broad range of energies and integration volumes.

In the present approach, we adopt the philosophy of IERM theory within RMT theory by allowing two electrons to escape into the outer region. Electron $i$ in the outer region is here described through a finite-difference (FD) representation for the radial coordinate $r_{i}$. We thus employ three distinct regions: (1) an inner region, in which all $N$ electrons are within a distance $a$ from the nucleus, represented with a standard R-matrix basis set, (2) a one-electron outer region, in which we combine a basis set representation for the residual system with $(N-1)$ electrons,  and a finite-difference representation for the ionised electron  with $r_N>a$, and (3) a two-electron outer region, in which we use a basis set representation for the residual system with $(N-2)$ electrons  and a finite-difference representation for the two electrons  with $r_{N-1}, r_N>a$.

As a proof-of-principle, we apply it in the present communication to single-photon double photoionization of He. This process has already been studied extensively, both experimentally \cite{Richter2009} and theoretically \cite{Kheifets1998,Meyer1997}, and it is therefore possible to compare the outcomes of the calculations with benchmark data. However, since the approach described here allows the simultaneous calculation of higher-order processes (such as double electron above threshold ionisation \cite{Parker2001}), we also expect signatures of these higher-order processes to appear in the results.

The Hamiltonian for the helium atom in a laser field is given in atomic units as
\begin{equation}
	\hat{H} = -\frac{1}{2}\nabla_1^2 -\frac{1}{2}\nabla_2^2-\frac{2}{r_1}-\frac{2}{r_2}+\frac{1}{r_{12}}+ E(t)\cdot(z_1+z_2)
\end{equation}
and the time-dependent Schr\"odinger equation is given by
\begin{equation}
i\frac{\partial\Psi}{\partial t} = \hat{H}\Psi.
\label{eq:tdse}
\end{equation}
In these equations, $r_{1}$ and $r_{2}$ are the radial coordinates of the first and second electron, and $\frac{1}{r_{12}}$ represents the inter electron repulsion. $E(t)$ is the time-dependent laser field, and $\Psi$ is the two-electron wavefunction. $z_1$ and $z_2$ are the positions of the electron in the direction of the laser field.

The two-electron outer region wavefunction is described on a set of two-dimensional FD grids given as $\Psi_{q}(r_{1},r_{2})$ where each grid describes the component of the wavefunction corresponding to the angular momentum quantum numbers q=($\ell_{1}, \ell_{2}, L$) across the coordinates $r_{1}=a...b$ and $r_{2}=a...b$. Near the boundary with the one-electron outer region ($r_{1} \approx a$ and $r_{2}\approx a$), the two-electron outer region is provided with the necessary wavefunction information from the one-electron outer region on an extension of the FD grid. This allows the direct evaluation of equation (\ref{eq:tdse}) using FD techniques. 

In the one-electron outer region, the inner electron is described using a near-complete basis of eigenfunctions of He$^+$ within a box with free-boundary conditions: eigenfunctions of
\begin{equation}
\hat{H}_{+} =  -\frac{1}{2}\nabla_1^2 -\frac{2}{r_1} +L_b \mathrm{\hspace{5pt}},
\end{equation}
where $L_b$ is a Bloch operator \cite{burke2011}. The outer electron is described using an FD representation. The presence of a Bloch operator for the inner electron leads to boundary derivative terms arising at the boundary between the one- and two-electron outer regions. As a consequence, the TDSE for the one electron outer region becomes

\begin{eqnarray}
i \frac{\partial}{\partial t} f_p(r_2,t) =  \hat{H}f_p(r_2,t) + \left. \frac{1}{2}\sum_{q}\mathcal{A}\Omega_{pq} \frac{\partial \Psi_{q}(r_1,r_2)}{\partial r_1} \right|_{r_1=a},
\end{eqnarray}
where $f_p(r_2,t)$ is an FD representation of the wavefunction at radial distance $r_{2}$ and time $t$ in channel $p$. $\Omega_{pq}$ is the surface amplitude that links channel $p$ to the two electron finite difference grid $\Psi_{q}(r_1,r_2)$, and $\mathcal{A} $ is the antisymmetrisation operator. This equation is similar to the propagation equation for inner region single ionisation RMT theory \cite{Nikolopoulos2008,Lysaght2012}. However, the final term on the right-hand side connects the wavefunction $f_p(r_2,t)$ in the one-electron outer region with the wavefunction $\Psi_{q}(r_1,r_2)$ in the two-electron outer region.

In the inner region, the propagation equations are identical to those in RMT theory for single ionisation \cite{Nikolopoulos2008,Lysaght2012}:
\begin{equation}
\frac{d}{dt}C_k(t) = -i\sum_{k'} H_{kk'}C_{k'}(t)-\frac{i}{2}\sum_p\omega_{pk} \left.\frac{\partial f_p(r_2,t)}{\partial r}\right |_{r=a}.
\end{equation}
In this equation, the coefficients $C_k(t)$ are coefficients of the two-electron R-matrix basis functions in the inner region, and $\omega_{pk}$ are boundary amplitudes of the two-electron functions at the boundary $r_2=a$.

RMT theory for single ionisation has previously adopted Arnoldi propagators for time propagation, requiring separate propagators for the homogeneous TDSE and for each boundary term that contributed to the propagation. The complexity arising from multiple propagators is avoided here through the use of a simpler Taylor series computational scheme.

At each stage of the calculation, we consider the anti-symmetry of the wavefunction under particle exchange. We have implemented this anti-symmetry by considering the two-electron outer region wavefunction for all $r_1 > a, r_2 > a$. The link between the one-electron outer region and two-electron outer region is taken along both boundaries of the two-electron outer region, $r_1 = a, r_2 \geq a$ and $r_1 \geq a, r_2 = a$. A phase change can be included to account for spin symmetry of the electron pair.

Analysis of the final wavefunction occurs through projection of the final state onto uncorrelated products of eigenfunctions of the one-electron He$^+$ hamiltonian:  $F_{\ell_{1}}(E_{1},r_{1})F_{\ell_{2}}(E_{2},r_{2})$.  As the two electrons travel far from the core and from each other, the energies can be interpreted as momenta $\mathbf{p}$ squared: $E_{i}=\mathbf{p_{i}}\cdot \mathbf{p_{i}}/2m=k_{i}^{2}\hbar^{2}/2m$, or in a.u.: $k_{i}=\sqrt{2E_{i}}$. In this communication we use $k$ rather than $E$ and generally refer to it as the momentum of the electron. 

This approach is correct in the limit in which the two ejected electrons are sufficiently weakly interacting. In the present calculations we propagate the electrons for (typically) 70 field periods after the laser pulse has ramped to zero, during which time their Coulomb repulsion enhances their spatial separation. We note that $k$ is derived from energy and includes contributions from the Coulomb potential, but these contributions appear to be negligible. If we allow the two-electron wave packet to depart the core for durations longer than the usual 70 field periods, then the calculated energy spectra do not change.

To separate the single-photon yield from the two-photon process, we define a cut-off $c$ such that all wavefunction population of momenta $k_{1},k_{2}$ where $k_{1}^{2}+k_{2}^{2}<c^{2}$ is considered to be resulting from the single-photon process. Here $c$ is chosen for an individual photon energy $E$ as $c>>\sqrt{2(E-I_{p})}$, where $I_{p}$ is the ionisation potential for He. From this we find a single-photon double-ionisation yield $\alpha$ from which we can calculate a corresponding cross section $\sigma^{2+}$ as $\sigma^{2+}=\alpha \omega /\int_{0}^{\infty}I(t)\mathrm{d}t$ ($I(t)$ is the intensity of the pulse at time $t$ and $\omega$ is the frequency of the pulse).

The presence of excited bound states (both single-electron and double-electron) in the final state wavefunction complicates the analysis of the final-state energy distributions of the two ejected electrons. As we are currently only interested in double-ionisation wave packets, a Gaussian mask was applied to the wavefunction in the regions $r_{1}<35\mathrm{a_{0}}$, $r_{2}<35\mathrm{a_{0}}$ to hide the singly bound states and $r_{1}^{2}+r_{2}^{2}<(150\mathrm{a_{0}})^{2}$ for the doubly-bound states. Care was taken to ensure that by the end of the time-propagation, the final yield had stabilised with respect to time, indicating that all double-ionisation wave packets had propagated into the unmasked area.

To demonstrate the capabilities of this method, five sets of photo-ionisation data were calculated for pulses of photon energies 84 eV, 99 eV, 125 eV, 150 eV, and 180 eV and of peak intensity $4 \times 10^{14}  \text{W/cm}^{2}$. The pulse is comprised of 15 cycles (two cycles of $\sin^{2}$ turn-on,  11 cycles of constant peak intensity, and two cycles of $\sin^{2}$ turn-off). A grid spacing of $\Delta x=0.25\mathrm{a_{0}}$ is used for both dimensions of the FD grid, and the boundary between the regions is placed at $b=25 \mathrm{a_{0}}$. The propagator is a 6th order Taylor series propagator with a time step of $0.028 \mathrm{as}$. A basis of 50 B-splines is used to describe the single electron functions in the inner region from which the two electron states were constructed. Two electron states with energies above 1000 a.u. were excluded from the calculations. Individual electrons are limited to a maximum angular momentum $\ell=3$ and the atom to a maximum angular momentum of $L=2$. The two-electron outer region is limited to $r_{1}+r_{2} \lesssim 900 a_{0}$.

\begin{figure}
	\begin{center}
		\includegraphics[width=8.5cm,angle=0]{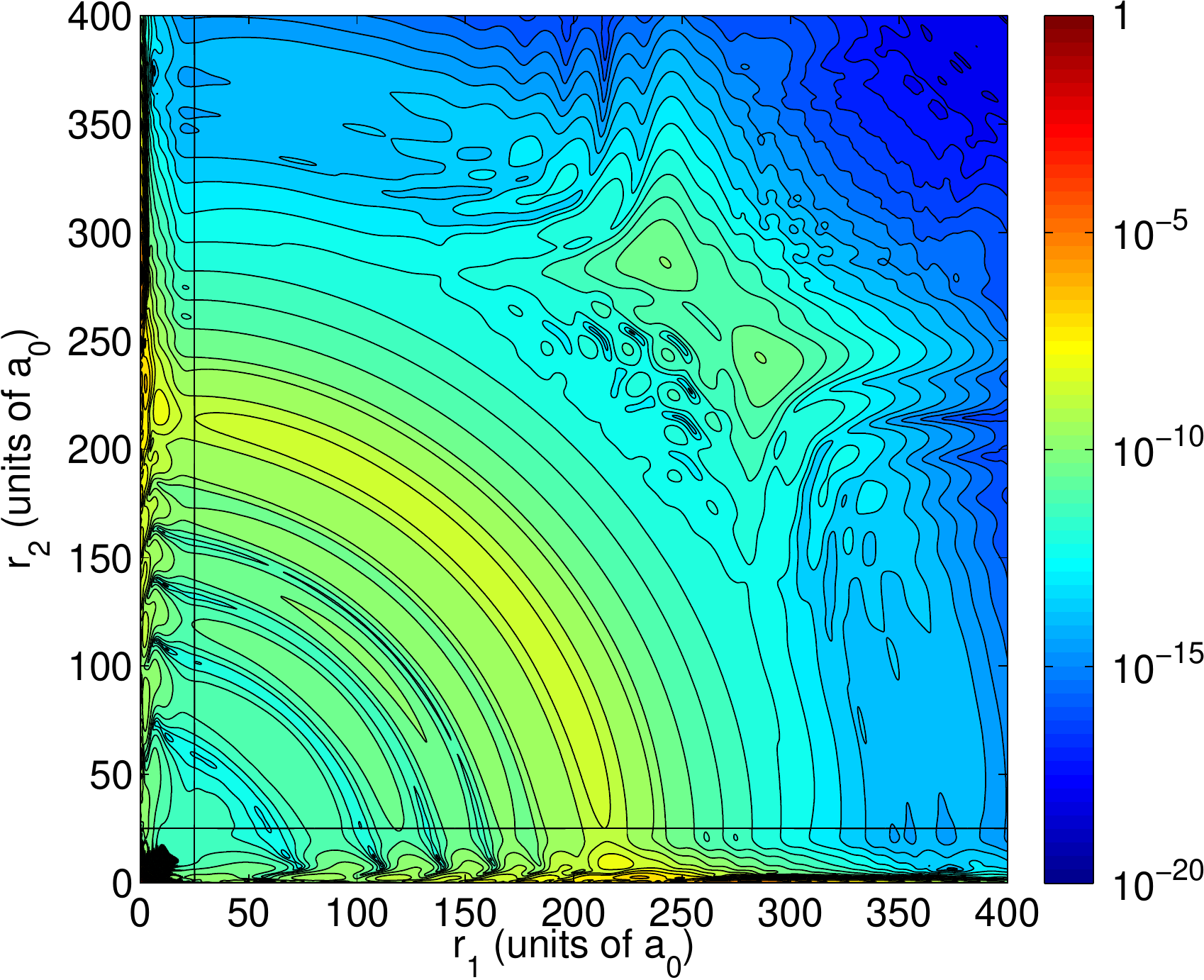}
		\caption{(Colour online) Two-electron helium wavefunction density ($\mathrm{a_{0}}^{-2}$) 70 cycles after the end of a 15 cycle pulse. $r_{1}$ corresponds to the radial distance of electron 1 and $r_{2}$ corresponds to the radial distance of electron 2. The pulse has a photon energy of 150 eV and a peak intensity of $10^{14}$ W cm$^{-2}$. All distances are given in bohr radii ($\mathrm{a_{0}}$). Boundaries at $r_{1},r_{2}=25\mathrm{a_{0}}$ divide the three RMT regions.}
		\label{WavefnRun}
	\end{center}
\end{figure}

\begin{figure}
	\begin{center}		\includegraphics[width=8.5cm,angle=0]{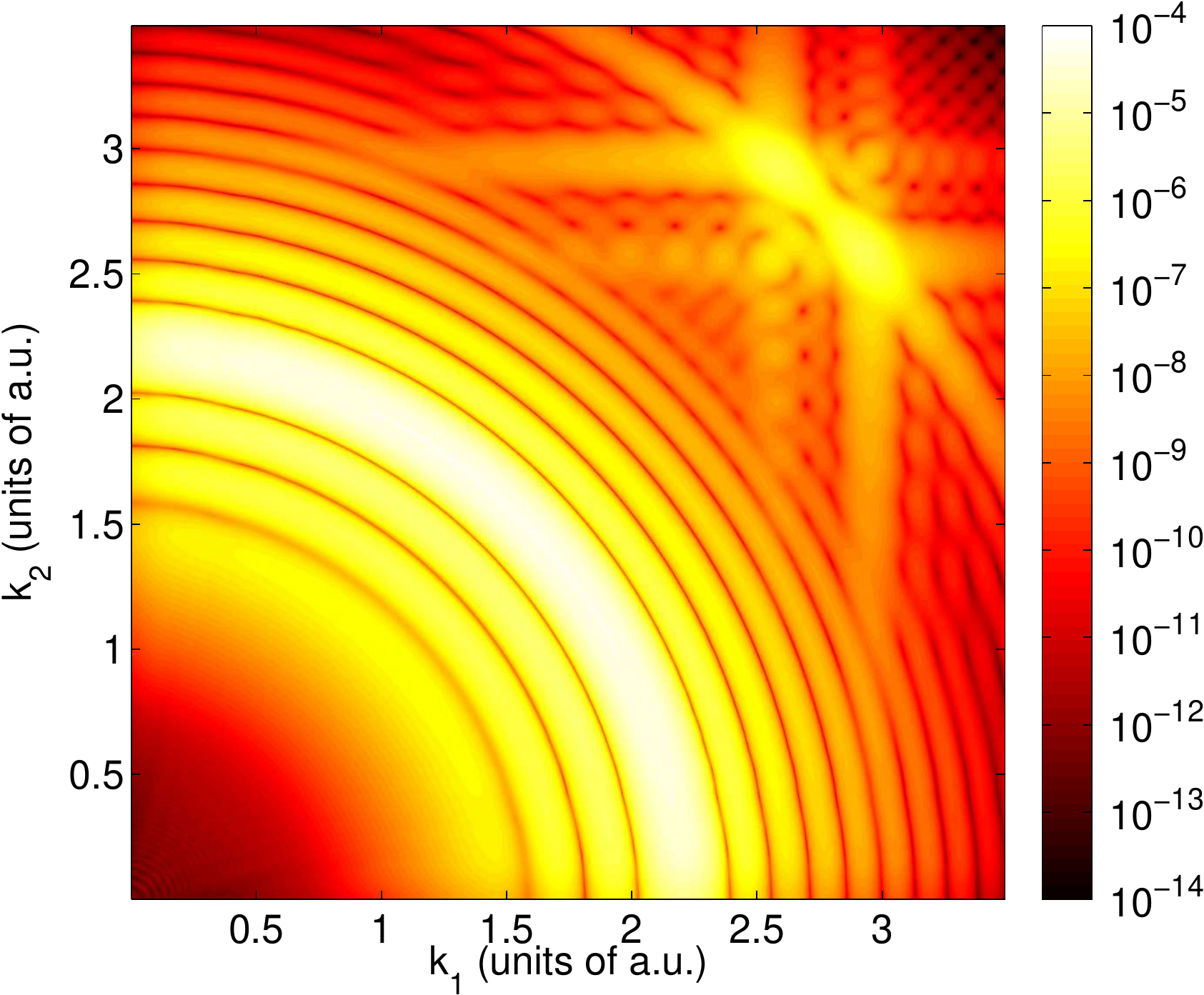}
		\caption{(Colour online) Two-electron probability distribution in momentum space calculated from the final state wavefunction (shown in figure \ref{WavefnRun}).  $k_{1}$ corresponds to the radial momentum of electron 1 and $k_{2}$ corresponds to the radial momentum of electron 2. The pulse has a photon energy of 150 eV and a peak intensity of $10^{14}$ W cm$^{-2}$. All momenta are given in $\mathrm{a.u.}$.}
		\label{ExampleMomentum}
	\end{center}
\end{figure}

Figure \ref{WavefnRun} shows a two electron wavefunction 70 cycles after the end of a 15 cycle laser pulse of photon energy 150 eV with the momentum transform for the large $t$ limit calculated from this wavefunction given in figure \ref{ExampleMomentum}. A single-photon non-sequential process (where both electrons are simultaneously ionised by a single photon) is indicated in the momentum plot by the arc at $(k_{1}^{2}+k_{2}^{2})\approx(2.3\mathrm{a.u.})^{2}$, as would be predicted given the necessary energy sharing for this process. This process is found in the corresponding wavefunction density (figure \ref{WavefnRun}) in the arc in the region $r_{1}^{2}+r_{2}^{2}=(220\mathrm{a_{0}})^{2}$. In addition, evidence of a two-photon sequential process is seen in figure \ref{WavefnRun}, with the outer and inner electrons at a distance of $\approx 300\mathrm{a_{0}}$ and  $\approx 250\mathrm{a_{0}}$ from the nucleus respectively. This process may also be seen in the momentum transform for $k_{1}$ and $k_{2} \approx 3\mathrm{a.u.}$.

Processes other than double ionisation are also calculated using this approach, and their effects are visible in figure \ref{WavefnRun}. Single ionisation is visible close to axes $r_{1}, r_{2}=0$, and excitation to doubly-excited states is visible near the nucleus, in addition to the remaining population in the ground state. While further information about these processes can in theory be extracted from the final wavefunction data, none of these processes appear in the momentum transform due to the Gaussian mask acting on all non double-ionisation processes.

\begin{figure}
	\begin{center}
		\includegraphics[height=8.5cm,angle=270]{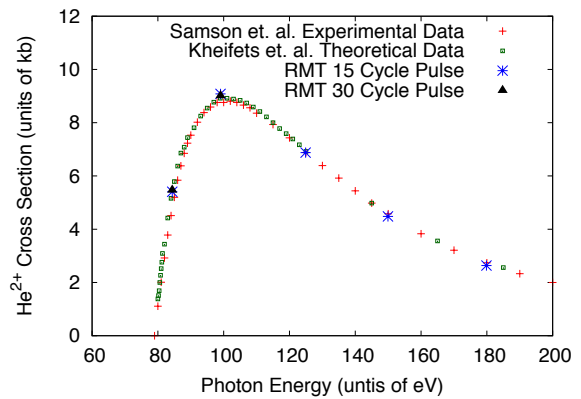}
		\caption{(Colour online) Single-photon double ionisation cross sections calculated using RMT alongside experimental data from \cite{Samson1998} and theoretical data from \cite{Kheifets1996}.} 
		\label{CrossSections}
	\end{center}
\end{figure}

Figure \ref{CrossSections} shows single photon double-ionisation cross sections obtained from a yield calculated using the momentum transform of the final wavefunctions. This data is given in comparison with experimental data \cite{Samson1998} and theoretical data \cite{Kheifets1996}. The theoretical RMT data follows the overall pattern of the experimental data, with the greatest disagreement ($\approx 15\%$) seen at 84 eV, reducing to $\approx 4\%$ at 125 eV, 150 eV and 180 eV. To examine the effect of pulse length, cross-sections were calculated at photon energies of 84 eV and 99 eV with a 30 cycle pulse (4 cycles ramp-on and 4 cycles ramp-off). Similar cross-sections were calculated for both pulse lengths, indicating that the cause of the greater error at these photon energies lies elsewhere. 

At low momenta, the double ionised He$^{2+}$ wave packet is difficult to distinguish from the long tails of higher energy He$^{+}$ bound states. The calculation of the final double ionisation yield for these near-threshold photon energies most likely contains a contribution from these bound states. Since it is difficult to measure this contribution exactly without propagating over a prohibitively large configuration space, we consider this effect to be a probable source of error. In addition, the differences between the current and the benchmark cross-sections may be reduced by extending the range of angular momenta over which the wave function is represented.

\begin{figure}
	\begin{center}
		\includegraphics[height=8.5cm,angle=270]{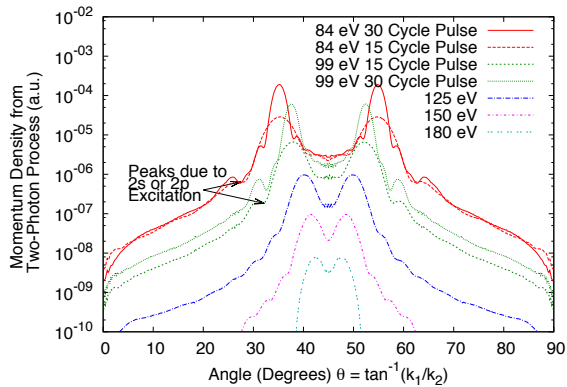}
		\caption{(Colour online) Two photon momentum density across an arc corresponding to $c_{min}^{2} < k_{1}^{2}+k_{2}^{2} < c_{max}^{2}$ where $k_{1}$ is the radial momentum of electron 1 and $k_{2}$ is the radial momentum of electron 2. Arrows indicating the expected angle of peaks caused by a sequential process involving an excitation of the un-ionised electron to 2s or 2p are shown for the 84 eV and 99 eV data.} 
		\label{2Sprocess}
	\end{center}
\end{figure}

To demonstrate that the RMT approach has the capacity to extract information about the two-photon sequential process (as well as the single photon nonsequential process), a mask was applied to the momentum transform so that only the momentum density in the region $c_{min}^{2} < k_{1}^{2}+k_{2}^{2} < c_{max}^{2}$ for the of angular momentum couplings $L=0$ and $L=2$ was retained. $c_{min}$ and $c_{max}$ are given values according to the photon energy for the two-photon sequential double ionisation momenta (for example for a photon energy of 150 eV in Figure \ref{ExampleMomentum}, $c_{min}=3.9$a.u. and $c_{max}=4.2$a.u.). The momentum density of the remaining arc is plotted against the angle $\theta=\tan^{-1}(k_{1}/k_{2})$ in figure \ref{2Sprocess}. These spectra are related to ejected-electron spectra, originally explored for sequential double ionisation by \cite{Laulan2003}, who showed their dependence on pulse length.

The two largest peaks in each photon energy are directly in the region expected for the sequential process where an electron is excited from the ground state into the continuum, leaving the bound electron in the He$^+$ 1s state which is ionised by a later photon. In the data for 84 eV and 99 eV features an order of magnitude smaller are visible, which occur at momenta corresponding to a two photon process (shown by the arrows in figure \ref{2Sprocess}) where the bound He$^+$ electron is excited to either the 2s or 2p state before being ionised by the second photon (as discussed in detail in \cite{Horner2007, Feist2009}). While these processes should also be present in the 125 eV, 150 eV and 180 eV spectra, they occur at angles where the 1s process dominates, making them difficult to observe. The 30 cycle pulse data shows these features more distinctly than the 15 cycle pulse. For these longer pulses, an additional minimum is seen in the 84 eV spectra, corresponding to a side-band caused by the pulse length.

In conclusion, we have combined a two-electron FD outer-region approach with the RMT method. This allows the double-ionised wave-packet to be propagated over a larger configuration space. The accuracy of the approach is demonstrated by the determination of He single-photon double ionisation cross sections for photon energies in the region 80 eV to 180 eV. We obtain agreement with experiment and existing theory to within 15\% near the single photon double ionisation threshold, and to within 4\% for higher photon energies. The capability to investigate higher-order processes is demonstrated through the observation of signatures associated with a sequential ionisation process involving excited states of the intermediate He$^{+}$ ion.

This agreement demonstrates the feasibility of attaching a two-electron FD region to an R-Matrix outer region using the RMT methods, and that this approach can be applied to predict a wide range of experimental observables. The FD method is highly parallelisable, and the current program scales linearly with the area covered by the two-electron double outer region.

The RMT method has the potential to study double-ionisation with full-correlation in general atoms. In order for this potential to be realised, it will be necessary to develop a multi-electron inner region basis set with full correlation to model double ionisation in the inner region, in addition to a corresponding outer region basis set.  

This research was sponsored by the Engineering and Physical Sciences Research Council (UK) under grant ref. no. G/055416/1 and by the Initial Training Network CORINF under a Marie Curie Action of the European Commission. This work used the ARCHER UK National Supercomputing service (http://www.archer.ac.uk). Data from figure \ref{WavefnRun} to figure \ref{2Sprocess} can be accessed via http://pure.qub.ac.uk/portal/en/datasets/search.html.

	\bibliography{Write}

\begin{thebibliography}{26}%
\makeatletter
\providecommand \@ifxundefined [1]{%
 \@ifx{#1\undefined}
}%
\providecommand \@ifnum [1]{%
 \ifnum #1\expandafter \@firstoftwo
 \else \expandafter \@secondoftwo
 \fi
}%
\providecommand \@ifx [1]{%
 \ifx #1\expandafter \@firstoftwo
 \else \expandafter \@secondoftwo
 \fi
}%
\providecommand \natexlab [1]{#1}%
\providecommand \enquote  [1]{``#1''}%
\providecommand \bibnamefont  [1]{#1}%
\providecommand \bibfnamefont [1]{#1}%
\providecommand \citenamefont [1]{#1}%
\providecommand \href@noop [0]{\@secondoftwo}%
\providecommand \href [0]{\begingroup \@sanitize@url \@href}%
\providecommand \@href[1]{\@@startlink{#1}\@@href}%
\providecommand \@@href[1]{\endgroup#1\@@endlink}%
\providecommand \@sanitize@url [0]{\catcode `\\12\catcode `\$12\catcode
  `\&12\catcode `\#12\catcode `\^12\catcode `\_12\catcode `\%12\relax}%
\providecommand \@@startlink[1]{}%
\providecommand \@@endlink[0]{}%
\providecommand \url  [0]{\begingroup\@sanitize@url \@url }%
\providecommand \@url [1]{\endgroup\@href {#1}{\urlprefix }}%
\providecommand \urlprefix  [0]{URL }%
\providecommand \Eprint [0]{\href }%
\providecommand \doibase [0]{http://dx.doi.org/}%
\providecommand \selectlanguage [0]{\@gobble}%
\providecommand \bibinfo  [0]{\@secondoftwo}%
\providecommand \bibfield  [0]{\@secondoftwo}%
\providecommand \translation [1]{[#1]}%
\providecommand \BibitemOpen [0]{}%
\providecommand \bibitemStop [0]{}%
\providecommand \bibitemNoStop [0]{.\EOS\space}%
\providecommand \EOS [0]{\spacefactor3000\relax}%
\providecommand \BibitemShut  [1]{\csname bibitem#1\endcsname}%
\let\auto@bib@innerbib\@empty
\bibitem [{\citenamefont {Corkum}\ and\ \citenamefont
  {Krausz}(2007)}]{Corkum2007}%
  \BibitemOpen
  \bibfield  {author} {\bibinfo {author} {\bibfnamefont {P.~B.}\ \bibnamefont
  {Corkum}}\ and\ \bibinfo {author} {\bibfnamefont {F.}~\bibnamefont
  {Krausz}},\ }\href {http://dx.doi.org/10.1038/nphys620} {\bibfield  {journal}
  {\bibinfo  {journal} {Nat Phys}\ }\textbf {\bibinfo {volume} {3}},\ \bibinfo
  {pages} {381} (\bibinfo {year} {2007})}\BibitemShut {NoStop}%
\bibitem [{\citenamefont {Krausz}\ and\ \citenamefont
  {Ivanov}(2009)}]{Krausz2009}%
  \BibitemOpen
  \bibfield  {author} {\bibinfo {author} {\bibfnamefont {F.}~\bibnamefont
  {Krausz}}\ and\ \bibinfo {author} {\bibfnamefont {M.}~\bibnamefont
  {Ivanov}},\ }\href {\doibase 10.1103/RevModPhys.81.163} {\bibfield  {journal}
  {\bibinfo  {journal} {Rev. Mod. Phys.}\ }\textbf {\bibinfo {volume} {81}},\
  \bibinfo {pages} {163} (\bibinfo {year} {2009})}\BibitemShut {NoStop}%
\bibitem [{\citenamefont {Yabashi}\ \emph {et~al.}(2013)\citenamefont
  {Yabashi}, \citenamefont {Tanaka}, \citenamefont {Tanaka}, \citenamefont
  {Tomizawa}, \citenamefont {Togashi}, \citenamefont {Nagasono}, \citenamefont
  {Ishikawa}, \citenamefont {Harries}, \citenamefont {Hikosaka}, \citenamefont
  {Hishikawa} \emph {et~al.}}]{Yabashi2013}%
  \BibitemOpen
  \bibfield  {author} {\bibinfo {author} {\bibfnamefont {M.}~\bibnamefont
  {Yabashi}}, \bibinfo {author} {\bibfnamefont {H.}~\bibnamefont {Tanaka}},
  \bibinfo {author} {\bibfnamefont {T.}~\bibnamefont {Tanaka}}, \bibinfo
  {author} {\bibfnamefont {H.}~\bibnamefont {Tomizawa}}, \bibinfo {author}
  {\bibfnamefont {T.}~\bibnamefont {Togashi}}, \bibinfo {author} {\bibfnamefont
  {M.}~\bibnamefont {Nagasono}}, \bibinfo {author} {\bibfnamefont
  {T.}~\bibnamefont {Ishikawa}}, \bibinfo {author} {\bibfnamefont
  {J.}~\bibnamefont {Harries}}, \bibinfo {author} {\bibfnamefont
  {Y.}~\bibnamefont {Hikosaka}}, \bibinfo {author} {\bibfnamefont
  {A.}~\bibnamefont {Hishikawa}},  \emph {et~al.},\ }\href@noop {} {\bibfield
  {journal} {\bibinfo  {journal} {Journal of Physics B: Atomic, Molecular and
  Optical Physics}\ }\textbf {\bibinfo {volume} {46}},\ \bibinfo {pages}
  {164001} (\bibinfo {year} {2013})}\BibitemShut {NoStop}%
\bibitem [{\citenamefont {Bergues}\ \emph {et~al.}(2012)\citenamefont
  {Bergues}, \citenamefont {K{\"u}bel}, \citenamefont {Johnson}, \citenamefont
  {Fischer}, \citenamefont {Camus}, \citenamefont {Betsch}, \citenamefont
  {Herrwerth}, \citenamefont {Senftleben}, \citenamefont {Sayler},
  \citenamefont {Rathje}, \citenamefont {Pfeifer}, \citenamefont {Ben-Itzhak},
  \citenamefont {Jones}, \citenamefont {Paulus}, \citenamefont {Krausz},
  \citenamefont {Moshammer}, \citenamefont {Ullrich},\ and\ \citenamefont
  {Kling}}]{Bergues2012}%
  \BibitemOpen
  \bibfield  {author} {\bibinfo {author} {\bibfnamefont {B.}~\bibnamefont
  {Bergues}}, \bibinfo {author} {\bibfnamefont {M.}~\bibnamefont {K{\"u}bel}},
  \bibinfo {author} {\bibfnamefont {N.~G.}\ \bibnamefont {Johnson}}, \bibinfo
  {author} {\bibfnamefont {B.}~\bibnamefont {Fischer}}, \bibinfo {author}
  {\bibfnamefont {N.}~\bibnamefont {Camus}}, \bibinfo {author} {\bibfnamefont
  {K.~J.}\ \bibnamefont {Betsch}}, \bibinfo {author} {\bibfnamefont
  {O.}~\bibnamefont {Herrwerth}}, \bibinfo {author} {\bibfnamefont
  {A.}~\bibnamefont {Senftleben}}, \bibinfo {author} {\bibfnamefont {A.~M.}\
  \bibnamefont {Sayler}}, \bibinfo {author} {\bibfnamefont {T.}~\bibnamefont
  {Rathje}}, \bibinfo {author} {\bibfnamefont {T.}~\bibnamefont {Pfeifer}},
  \bibinfo {author} {\bibfnamefont {I.}~\bibnamefont {Ben-Itzhak}}, \bibinfo
  {author} {\bibfnamefont {R.~R.}\ \bibnamefont {Jones}}, \bibinfo {author}
  {\bibfnamefont {G.~G.}\ \bibnamefont {Paulus}}, \bibinfo {author}
  {\bibfnamefont {F.}~\bibnamefont {Krausz}}, \bibinfo {author} {\bibfnamefont
  {R.}~\bibnamefont {Moshammer}}, \bibinfo {author} {\bibfnamefont
  {J.}~\bibnamefont {Ullrich}}, \ and\ \bibinfo {author} {\bibfnamefont
  {M.~F.}\ \bibnamefont {Kling}},\ }\href
  {http://dx.doi.org/10.1038/ncomms1807} {\bibfield  {journal} {\bibinfo
  {journal} {Nat Commun}\ }\textbf {\bibinfo {volume} {3}},\ \bibinfo {pages}
  {813} (\bibinfo {year} {2012})}\BibitemShut {NoStop}%
\bibitem [{\citenamefont {Yip}\ \emph {et~al.}(2013)\citenamefont {Yip},
  \citenamefont {Rescigno}, \citenamefont {McCurdy},\ and\ \citenamefont
  {Mart\'in}}]{Yip2013}%
  \BibitemOpen
  \bibfield  {author} {\bibinfo {author} {\bibfnamefont {F.~L.}\ \bibnamefont
  {Yip}}, \bibinfo {author} {\bibfnamefont {T.~N.}\ \bibnamefont {Rescigno}},
  \bibinfo {author} {\bibfnamefont {C.~W.}\ \bibnamefont {McCurdy}}, \ and\
  \bibinfo {author} {\bibfnamefont {F.}~\bibnamefont {Mart\'in}},\ }\href
  {\doibase 10.1103/PhysRevLett.110.173001} {\bibfield  {journal} {\bibinfo
  {journal} {Phys. Rev. Lett.}\ }\textbf {\bibinfo {volume} {110}},\ \bibinfo
  {pages} {173001} (\bibinfo {year} {2013})}\BibitemShut {NoStop}%
\bibitem [{\citenamefont {Bartschat}\ \emph {et~al.}(1996)\citenamefont
  {Bartschat}, \citenamefont {Hudson}, \citenamefont {Scott}, \citenamefont
  {Burke},\ and\ \citenamefont {Burke}}]{Bartschat1996}%
  \BibitemOpen
  \bibfield  {author} {\bibinfo {author} {\bibfnamefont {K.}~\bibnamefont
  {Bartschat}}, \bibinfo {author} {\bibfnamefont {E.~T.}\ \bibnamefont
  {Hudson}}, \bibinfo {author} {\bibfnamefont {M.~P.}\ \bibnamefont {Scott}},
  \bibinfo {author} {\bibfnamefont {P.~G.}\ \bibnamefont {Burke}}, \ and\
  \bibinfo {author} {\bibfnamefont {V.~M.}\ \bibnamefont {Burke}},\ }\href
  {http://stacks.iop.org/0953-4075/29/i=1/a=015} {\bibfield  {journal}
  {\bibinfo  {journal} {Journal of Physics B: Atomic, Molecular and Optical
  Physics}\ }\textbf {\bibinfo {volume} {29}},\ \bibinfo {pages} {115}
  (\bibinfo {year} {1996})}\BibitemShut {NoStop}%
\bibitem [{\citenamefont {Burke}\ \emph {et~al.}(1987)\citenamefont {Burke},
  \citenamefont {Noble},\ and\ \citenamefont {Scott}}]{Burke1987}%
  \BibitemOpen
  \bibfield  {author} {\bibinfo {author} {\bibfnamefont {P.~G.}\ \bibnamefont
  {Burke}}, \bibinfo {author} {\bibfnamefont {C.~J.}\ \bibnamefont {Noble}}, \
  and\ \bibinfo {author} {\bibfnamefont {P.}~\bibnamefont {Scott}},\ }\href
  {\doibase 10.1098/rspa.1987.0040} {\bibfield  {journal} {\bibinfo  {journal}
  {Proceedings of the Royal Society of London A: Mathematical, Physical and
  Engineering Sciences}\ }\textbf {\bibinfo {volume} {410}},\ \bibinfo {pages}
  {289} (\bibinfo {year} {1987})}\BibitemShut {NoStop}%
\bibitem [{\citenamefont {Scott}\ \emph {et~al.}(2009)\citenamefont {Scott},
  \citenamefont {Scott}, \citenamefont {Burke}, \citenamefont {Stitt},
  \citenamefont {Faro-Maza}, \citenamefont {Denis},\ and\ \citenamefont
  {Maniopoulou}}]{Scott2009}%
  \BibitemOpen
  \bibfield  {author} {\bibinfo {author} {\bibfnamefont {N.}~\bibnamefont
  {Scott}}, \bibinfo {author} {\bibfnamefont {M.}~\bibnamefont {Scott}},
  \bibinfo {author} {\bibfnamefont {P.}~\bibnamefont {Burke}}, \bibinfo
  {author} {\bibfnamefont {T.}~\bibnamefont {Stitt}}, \bibinfo {author}
  {\bibfnamefont {V.}~\bibnamefont {Faro-Maza}}, \bibinfo {author}
  {\bibfnamefont {C.}~\bibnamefont {Denis}}, \ and\ \bibinfo {author}
  {\bibfnamefont {A.}~\bibnamefont {Maniopoulou}},\ }\href {\doibase
  http://dx.doi.org/10.1016/j.cpc.2009.07.018} {\bibfield  {journal} {\bibinfo
  {journal} {Computer Physics Communications}\ }\textbf {\bibinfo {volume}
  {180}},\ \bibinfo {pages} {2424 } (\bibinfo {year} {2009})}\BibitemShut
  {NoStop}%
\bibitem [{\citenamefont {McIntyre}\ \emph {et~al.}(2013)\citenamefont
  {McIntyre}, \citenamefont {Kinnen},\ and\ \citenamefont
  {Scott}}]{McIntyre2013}%
  \BibitemOpen
  \bibfield  {author} {\bibinfo {author} {\bibfnamefont {M.~W.}\ \bibnamefont
  {McIntyre}}, \bibinfo {author} {\bibfnamefont {A.~J.}\ \bibnamefont
  {Kinnen}}, \ and\ \bibinfo {author} {\bibfnamefont {M.~P.}\ \bibnamefont
  {Scott}},\ }\href {\doibase 10.1103/PhysRevA.88.053413} {\bibfield  {journal}
  {\bibinfo  {journal} {Phys. Rev. A}\ }\textbf {\bibinfo {volume} {88}},\
  \bibinfo {pages} {053413} (\bibinfo {year} {2013})}\BibitemShut {NoStop}%
\bibitem [{\citenamefont {Moore}\ \emph {et~al.}(2011)\citenamefont {Moore},
  \citenamefont {Lysaght}, \citenamefont {Nikolopoulos}, \citenamefont
  {Parker}, \citenamefont {{Van Der Hart}},\ and\ \citenamefont
  {Taylor}}]{Moore2011}%
  \BibitemOpen
  \bibfield  {author} {\bibinfo {author} {\bibfnamefont {L.}~\bibnamefont
  {Moore}}, \bibinfo {author} {\bibfnamefont {M.}~\bibnamefont {Lysaght}},
  \bibinfo {author} {\bibfnamefont {L.}~\bibnamefont {Nikolopoulos}}, \bibinfo
  {author} {\bibfnamefont {J.}~\bibnamefont {Parker}}, \bibinfo {author}
  {\bibfnamefont {H.}~\bibnamefont {{Van Der Hart}}}, \ and\ \bibinfo {author}
  {\bibfnamefont {K.}~\bibnamefont {Taylor}},\ }\href {\doibase
  10.1080/09500340.2011.559315} {\bibfield  {journal} {\bibinfo  {journal}
  {Journal of Modern Optics}\ }\textbf {\bibinfo {volume} {58}},\ \bibinfo
  {pages} {1132} (\bibinfo {year} {2011})}\BibitemShut {NoStop}%
\bibitem [{\citenamefont {Lysaght}\ \emph {et~al.}(2012)\citenamefont
  {Lysaght}, \citenamefont {Moore}, \citenamefont {Nikolopoulos}, \citenamefont
  {Parker}, \citenamefont {van~der Hart},\ and\ \citenamefont
  {Taylor}}]{Lysaght2012}%
  \BibitemOpen
  \bibfield  {author} {\bibinfo {author} {\bibfnamefont {M.~A.}\ \bibnamefont
  {Lysaght}}, \bibinfo {author} {\bibfnamefont {L.~R.}\ \bibnamefont {Moore}},
  \bibinfo {author} {\bibfnamefont {L.~A.~A.}\ \bibnamefont {Nikolopoulos}},
  \bibinfo {author} {\bibfnamefont {J.~S.}\ \bibnamefont {Parker}}, \bibinfo
  {author} {\bibfnamefont {H.~W.}\ \bibnamefont {van~der Hart}}, \ and\
  \bibinfo {author} {\bibfnamefont {K.~T.}\ \bibnamefont {Taylor}},\ }\href
  {http://stacks.iop.org/1742-6596/388/i=1/a=012027} {\bibfield  {journal}
  {\bibinfo  {journal} {Journal of Physics: Conference Series}\ }\textbf
  {\bibinfo {volume} {388}},\ \bibinfo {pages} {012027} (\bibinfo {year}
  {2012})}\BibitemShut {NoStop}%
\bibitem [{\citenamefont {Nikolopoulos}\ \emph {et~al.}(2008)\citenamefont
  {Nikolopoulos}, \citenamefont {Parker},\ and\ \citenamefont
  {Taylor}}]{Nikolopoulos2008}%
  \BibitemOpen
  \bibfield  {author} {\bibinfo {author} {\bibfnamefont {L.~A.~A.}\
  \bibnamefont {Nikolopoulos}}, \bibinfo {author} {\bibfnamefont {J.~S.}\
  \bibnamefont {Parker}}, \ and\ \bibinfo {author} {\bibfnamefont {K.~T.}\
  \bibnamefont {Taylor}},\ }\href {\doibase 10.1103/PhysRevA.78.063420}
  {\bibfield  {journal} {\bibinfo  {journal} {Phys. Rev. A}\ }\textbf {\bibinfo
  {volume} {78}},\ \bibinfo {pages} {063420} (\bibinfo {year}
  {2008})}\BibitemShut {NoStop}%
\bibitem [{\citenamefont {Rey}\ and\ \citenamefont {van~der
  Hart}(2014)}]{Rey2014}%
  \BibitemOpen
  \bibfield  {author} {\bibinfo {author} {\bibfnamefont {H.~F.}\ \bibnamefont
  {Rey}}\ and\ \bibinfo {author} {\bibfnamefont {H.~W.}\ \bibnamefont {van~der
  Hart}},\ }\href {\doibase 10.1103/PhysRevA.90.033402} {\bibfield  {journal}
  {\bibinfo  {journal} {Phys. Rev. A}\ }\textbf {\bibinfo {volume} {90}},\
  \bibinfo {pages} {033402} (\bibinfo {year} {2014})}\BibitemShut {NoStop}%
\bibitem [{\citenamefont {van~der Hart}\ and\ \citenamefont
  {Morgan}(2014)}]{Morgan2014}%
  \BibitemOpen
  \bibfield  {author} {\bibinfo {author} {\bibfnamefont {H.~W.}\ \bibnamefont
  {van~der Hart}}\ and\ \bibinfo {author} {\bibfnamefont {R.}~\bibnamefont
  {Morgan}},\ }\href {\doibase 10.1103/PhysRevA.90.013424} {\bibfield
  {journal} {\bibinfo  {journal} {Phys. Rev. A}\ }\textbf {\bibinfo {volume}
  {90}},\ \bibinfo {pages} {013424} (\bibinfo {year} {2014})}\BibitemShut
  {NoStop}%
\bibitem [{\citenamefont {Burke}(2011)}]{burke2011}%
  \BibitemOpen
  \bibfield  {author} {\bibinfo {author} {\bibfnamefont {P.}~\bibnamefont
  {Burke}},\ }\href {http://books.google.co.uk/books?id=B-jQiI8SQY0C} {\emph
  {\bibinfo {title} {R-Matrix Theory of Atomic Collisions: Application to
  Atomic, Molecular and Optical Processes}}},\ Springer Series on Atomic,
  Optical, and Plasma Physics\ (\bibinfo  {publisher} {Springer},\ \bibinfo
  {year} {2011})\BibitemShut {NoStop}%
\bibitem [{\citenamefont {Smyth}\ \emph {et~al.}(1998)\citenamefont {Smyth},
  \citenamefont {Parker},\ and\ \citenamefont {Taylor}}]{Smyth1998}%
  \BibitemOpen
  \bibfield  {author} {\bibinfo {author} {\bibfnamefont {E.~S.}\ \bibnamefont
  {Smyth}}, \bibinfo {author} {\bibfnamefont {J.~S.}\ \bibnamefont {Parker}}, \
  and\ \bibinfo {author} {\bibfnamefont {K.}~\bibnamefont {Taylor}},\ }\href
  {\doibase http://dx.doi.org/10.1016/S0010-4655(98)00083-6} {\bibfield
  {journal} {\bibinfo  {journal} {Computer Physics Communications}\ }\textbf
  {\bibinfo {volume} {114}},\ \bibinfo {pages} {1 } (\bibinfo {year}
  {1998})}\BibitemShut {NoStop}%
\bibitem [{\citenamefont {van~der Hart}(2014)}]{Hugo2014}%
  \BibitemOpen
  \bibfield  {author} {\bibinfo {author} {\bibfnamefont {H.~W.}\ \bibnamefont
  {van~der Hart}},\ }\href {\doibase 10.1103/PhysRevA.89.053407} {\bibfield
  {journal} {\bibinfo  {journal} {Phys. Rev. A}\ }\textbf {\bibinfo {volume}
  {89}},\ \bibinfo {pages} {053407} (\bibinfo {year} {2014})}\BibitemShut
  {NoStop}%
\bibitem [{\citenamefont {Richter}\ \emph {et~al.}(2009)\citenamefont
  {Richter}, \citenamefont {Amusia}, \citenamefont {Bobashev}, \citenamefont
  {Feigl}, \citenamefont {Jurani\ifmmode~\acute{c}\else \'{c}\fi{}},
  \citenamefont {Martins}, \citenamefont {Sorokin},\ and\ \citenamefont
  {Tiedtke}}]{Richter2009}%
  \BibitemOpen
  \bibfield  {author} {\bibinfo {author} {\bibfnamefont {M.}~\bibnamefont
  {Richter}}, \bibinfo {author} {\bibfnamefont {M.~Y.}\ \bibnamefont {Amusia}},
  \bibinfo {author} {\bibfnamefont {S.~V.}\ \bibnamefont {Bobashev}}, \bibinfo
  {author} {\bibfnamefont {T.}~\bibnamefont {Feigl}}, \bibinfo {author}
  {\bibfnamefont {P.~N.}\ \bibnamefont {Jurani\ifmmode~\acute{c}\else
  \'{c}\fi{}}}, \bibinfo {author} {\bibfnamefont {M.}~\bibnamefont {Martins}},
  \bibinfo {author} {\bibfnamefont {A.~A.}\ \bibnamefont {Sorokin}}, \ and\
  \bibinfo {author} {\bibfnamefont {K.}~\bibnamefont {Tiedtke}},\ }\href
  {\doibase 10.1103/PhysRevLett.102.163002} {\bibfield  {journal} {\bibinfo
  {journal} {Phys. Rev. Lett.}\ }\textbf {\bibinfo {volume} {102}},\ \bibinfo
  {pages} {163002} (\bibinfo {year} {2009})}\BibitemShut {NoStop}%
\bibitem [{\citenamefont {Kheifets}\ and\ \citenamefont
  {Bray}(1998)}]{Kheifets1998}%
  \BibitemOpen
  \bibfield  {author} {\bibinfo {author} {\bibfnamefont {A.~S.}\ \bibnamefont
  {Kheifets}}\ and\ \bibinfo {author} {\bibfnamefont {I.}~\bibnamefont
  {Bray}},\ }\href {http://stacks.iop.org/0953-4075/31/i=10/a=002} {\bibfield
  {journal} {\bibinfo  {journal} {Journal of Physics B: Atomic, Molecular and
  Optical Physics}\ }\textbf {\bibinfo {volume} {31}},\ \bibinfo {pages} {L447}
  (\bibinfo {year} {1998})}\BibitemShut {NoStop}%
\bibitem [{\citenamefont {Meyer}\ \emph {et~al.}(1997)\citenamefont {Meyer},
  \citenamefont {Greene},\ and\ \citenamefont {Esry}}]{Meyer1997}%
  \BibitemOpen
  \bibfield  {author} {\bibinfo {author} {\bibfnamefont {K.~W.}\ \bibnamefont
  {Meyer}}, \bibinfo {author} {\bibfnamefont {C.~H.}\ \bibnamefont {Greene}}, \
  and\ \bibinfo {author} {\bibfnamefont {B.~D.}\ \bibnamefont {Esry}},\ }\href
  {\doibase 10.1103/PhysRevLett.78.4902} {\bibfield  {journal} {\bibinfo
  {journal} {Phys. Rev. Lett.}\ }\textbf {\bibinfo {volume} {78}},\ \bibinfo
  {pages} {4902} (\bibinfo {year} {1997})}\BibitemShut {NoStop}%
\bibitem [{\citenamefont {Parker}\ \emph {et~al.}(2001)\citenamefont {Parker},
  \citenamefont {Moore}, \citenamefont {Meharg}, \citenamefont {Dundas},\ and\
  \citenamefont {Taylor}}]{Parker2001}%
  \BibitemOpen
  \bibfield  {author} {\bibinfo {author} {\bibfnamefont {J.~S.}\ \bibnamefont
  {Parker}}, \bibinfo {author} {\bibfnamefont {L.~R.}\ \bibnamefont {Moore}},
  \bibinfo {author} {\bibfnamefont {K.~J.}\ \bibnamefont {Meharg}}, \bibinfo
  {author} {\bibfnamefont {D.}~\bibnamefont {Dundas}}, \ and\ \bibinfo {author}
  {\bibfnamefont {K.~T.}\ \bibnamefont {Taylor}},\ }\href
  {http://stacks.iop.org/0953-4075/34/i=3/a=103} {\bibfield  {journal}
  {\bibinfo  {journal} {Journal of Physics B: Atomic, Molecular and Optical
  Physics}\ }\textbf {\bibinfo {volume} {34}},\ \bibinfo {pages} {L69}
  (\bibinfo {year} {2001})}\BibitemShut {NoStop}%
\bibitem [{\citenamefont {Samson}\ \emph {et~al.}(1998)\citenamefont {Samson},
  \citenamefont {Stolte}, \citenamefont {He}, \citenamefont {Cutler},
  \citenamefont {Lu},\ and\ \citenamefont {Bartlett}}]{Samson1998}%
  \BibitemOpen
  \bibfield  {author} {\bibinfo {author} {\bibfnamefont {J.~A.~R.}\
  \bibnamefont {Samson}}, \bibinfo {author} {\bibfnamefont {W.~C.}\
  \bibnamefont {Stolte}}, \bibinfo {author} {\bibfnamefont {Z.-X.}\
  \bibnamefont {He}}, \bibinfo {author} {\bibfnamefont {J.~N.}\ \bibnamefont
  {Cutler}}, \bibinfo {author} {\bibfnamefont {Y.}~\bibnamefont {Lu}}, \ and\
  \bibinfo {author} {\bibfnamefont {R.~J.}\ \bibnamefont {Bartlett}},\ }\href
  {\doibase 10.1103/PhysRevA.57.1906} {\bibfield  {journal} {\bibinfo
  {journal} {Phys. Rev. A}\ }\textbf {\bibinfo {volume} {57}},\ \bibinfo
  {pages} {1906} (\bibinfo {year} {1998})}\BibitemShut {NoStop}%
\bibitem [{\citenamefont {Kheifets}\ and\ \citenamefont
  {Bray}(1996)}]{Kheifets1996}%
  \BibitemOpen
  \bibfield  {author} {\bibinfo {author} {\bibfnamefont {A.~S.}\ \bibnamefont
  {Kheifets}}\ and\ \bibinfo {author} {\bibfnamefont {I.}~\bibnamefont
  {Bray}},\ }\href {\doibase 10.1103/PhysRevA.54.R995} {\bibfield  {journal}
  {\bibinfo  {journal} {Phys. Rev. A}\ }\textbf {\bibinfo {volume} {54}},\
  \bibinfo {pages} {R995} (\bibinfo {year} {1996})}\BibitemShut {NoStop}%
\bibitem [{\citenamefont {Laulan}\ and\ \citenamefont
  {Bachau}(2003)}]{Laulan2003}%
  \BibitemOpen
  \bibfield  {author} {\bibinfo {author} {\bibfnamefont {S.}~\bibnamefont
  {Laulan}}\ and\ \bibinfo {author} {\bibfnamefont {H.}~\bibnamefont
  {Bachau}},\ }\href {\doibase 10.1103/PhysRevA.68.013409} {\bibfield
  {journal} {\bibinfo  {journal} {Phys. Rev. A}\ }\textbf {\bibinfo {volume}
  {68}},\ \bibinfo {pages} {013409} (\bibinfo {year} {2003})}\BibitemShut
  {NoStop}%
\bibitem [{\citenamefont {Horner}\ \emph {et~al.}(2007)\citenamefont {Horner},
  \citenamefont {Morales}, \citenamefont {Rescigno}, \citenamefont
  {Mart\'{i}n},\ and\ \citenamefont {McCurdy}}]{Horner2007}%
  \BibitemOpen
  \bibfield  {author} {\bibinfo {author} {\bibfnamefont {D.~A.}\ \bibnamefont
  {Horner}}, \bibinfo {author} {\bibfnamefont {F.}~\bibnamefont {Morales}},
  \bibinfo {author} {\bibfnamefont {T.~N.}\ \bibnamefont {Rescigno}}, \bibinfo
  {author} {\bibfnamefont {F.}~\bibnamefont {Mart\'{i}n}}, \ and\ \bibinfo
  {author} {\bibfnamefont {C.~W.}\ \bibnamefont {McCurdy}},\ }\href {\doibase
  10.1103/PhysRevA.76.030701} {\bibfield  {journal} {\bibinfo  {journal} {Phys.
  Rev. A}\ }\textbf {\bibinfo {volume} {76}},\ \bibinfo {pages} {030701}
  (\bibinfo {year} {2007})}\BibitemShut {NoStop}%
\bibitem [{\citenamefont {Feist}\ \emph {et~al.}(2009)\citenamefont {Feist},
  \citenamefont {Pazourek}, \citenamefont {Nagele}, \citenamefont {Persson},
  \citenamefont {Schneider}, \citenamefont {Collins},\ and\ \citenamefont
  {Burgdörfer}}]{Feist2009}%
  \BibitemOpen
  \bibfield  {author} {\bibinfo {author} {\bibfnamefont {J.}~\bibnamefont
  {Feist}}, \bibinfo {author} {\bibfnamefont {R.}~\bibnamefont {Pazourek}},
  \bibinfo {author} {\bibfnamefont {S.}~\bibnamefont {Nagele}}, \bibinfo
  {author} {\bibfnamefont {E.}~\bibnamefont {Persson}}, \bibinfo {author}
  {\bibfnamefont {B.~I.}\ \bibnamefont {Schneider}}, \bibinfo {author}
  {\bibfnamefont {L.~A.}\ \bibnamefont {Collins}}, \ and\ \bibinfo {author}
  {\bibfnamefont {J.}~\bibnamefont {Burgdörfer}},\ }\href
  {http://stacks.iop.org/0953-4075/42/i=13/a=134014} {\bibfield  {journal}
  {\bibinfo  {journal} {Journal of Physics B: Atomic, Molecular and Optical
  Physics}\ }\textbf {\bibinfo {volume} {42}},\ \bibinfo {pages} {134014}
  (\bibinfo {year} {2009})}\BibitemShut {NoStop}%
\end{thebibliography}%
\end{document}